\documentclass[12pt]{article}
\usepackage{graphicx,graphics}
\usepackage{amsmath}
\usepackage{amsfonts}
\usepackage{amssymb}
\title{Exact solutions for the general fifth order KdV equation by the extended tanh method}

\author{Alvaro Salas \thanks{Department of Mathematics, Universidad de
Caldas, Department of Mathematics, Universidad Nacional de Colombia,
Manizales, Colombia. \emph{email} : asalash2002@yahoo.com}\\
Cesar A. G\'omez S \thanks{Universidad Nacional de Colombia,
Bogot\'a. \emph{email} : cagomezsi@unal.edu.co}
\\
Jos\'e Gonzalo Escobar Lugo \thanks{ Universidad Cooperativa de
Colombia \emph{email} : jogoel@gmail.com}}
\date{}
\begin {document}
\maketitle
\begin {abstract}
In this paper we show some exact solutions for the general fifth
order KdV equation $u_t+ \omega\,u_{xxxxx}+ \alpha\,uu_{xxx}+\beta
u_xu_{xx}+\gamma u^2u_x=0$. These solutions are obtained by the
extended tanh method.
\end{abstract}
\section{Introduction}
A large variety of physical, chemical, and biological phenomena is
governed by nonlinear evolution equations. The analytical study of
nonlinear partial differential equations was of great interest
during the last decades. Investigations of traveling wave solutions
of nonlinear equations play an important role in the study of
nonlinear physical phenomena. The importance of obtaining the exact
solutions, if available, of those nonlinear equations facilitates
the verification of numerical solvers and aids in the stability
analysis of solutions.\newline The nonlinear generalized KdV
equation of fifth order (fKdV equation) reads
\begin{equation}\label{eq00}
u_t+\omega\, u_{xxxxx}+\alpha\, uu_{xxx}+\beta\, u_xu_{xx}+\gamma
u^2u_x=0,
\end{equation}
where $\alpha$, $\beta$, $\gamma$ and $\omega$ are arbitrary real
parameters.

This equation describes motions of long waves in shallow water under
gravity and in a one-dimensional nonlinear lattice  and it is an
important mathematical model with wide applications in quantum
mechanics and nonlinear optics. Typical examples are widely used in
various fields such as solid state physics, plasma physics, fluid
physics and quantum field theory. A great deal of research work has
been invested during the past decades for the study of the fKdV
equation. The main goal of these studies was directed towards its
analytical and numerical solutions. Several different approaches,
such as Blackund transformation, a bilinear form, and a Lax pair,
have been used independently by which soliton and multi-soliton
solutions are obtained. Ablowitz et al. \cite{ablo} implemented the
inverse scattering transform method to handle the nonlinear
equations of physical significance where soliton solutions and
rational solutions were developed. \newline Some important
particular cases of Eq. (\ref{eq00}) are :
\begin{itemize}
\item Kaup-Kupershmidt equation (KK equation) \cite{kk1}\cite{kk2}\cite{kk3}\cite{kk4}\cite{kk5}
\begin{equation}\label{eq01}
u_t+u_{xxxxx}+10 uu_{xxx}+25u_xu_{xx}+20 u^2u_x=0.
\end{equation}
\item Sawada-Kotera equation (SK equation) \cite{sawada}\cite{salas}
\begin{equation}\label{eq02}
u_t+u_{xxxxx}+5 uu_{xxx}+5 u_xu_{xx}+5 u^2u_x=0.
\end{equation}
\item Caudrey-Dodd-Gibbon equation
\begin{equation}\label{eq03}
u_t+ u_{xxxxx}+ 30uu_{xxx}+30u_xu_{xx}+180u^2u_x=0.
\end{equation}
\item Lax equation \cite{lax}
\begin{equation}\label{eq04}
u_t+ u_{xxxxx}+\, 10 \,uu_{xxx}+20 u_xu_{xx}+30 u^2u_x=0.
\end{equation}
\item Ito equation \cite{ito}\cite{ito1}
\begin{equation}\label{eq05}
u_t+ u_{xxxxx}+3 \,uu_{xxx}+6 u_xu_{xx}+2 \,u^2u_x=0.
\end{equation}
\end{itemize}
As the constants $\alpha$, $\beta$ and $\gamma$
 change, the properties of the equation (\ref{eq00}) drastically change. For
instance, the Lax equation with $\alpha= 10$, $\beta = 20$, and
$\gamma = 30$, and the SK equation where $\alpha= = \beta =\gamma=
5$, are completely integrable. These two equations have $N$-soliton
solutions and an infinite set of conserved densities. Another
example is the KK equation with $\alpha = 10$, $\beta = 25$, and
$\gamma = 20$, which is known to be integrable \cite{kk3}, and has
bilinear representations \cite{kk3}\cite{kk5}, but for which the
explicit form of the $N$-soliton solutions is not known. A fourth
equation in this class is the Ito equation, with $\alpha = 3$,
$\beta = 6$, and $\gamma = 2$, which is not completely integrable,
but has a limited number of special conserved densities \cite{ito1}.

From 70's, a vast variety of simple and direct methods to find
analytic solutions of nonlinear differential equations and evolution
equations have been developed. Recently, the extended tanh method
\cite{abdou} has been successfully used in seeking the solitary wave
solution and other kinds of solutions.

We will find solutions of Eq. (\ref{eq00}) for $\omega\neq 0$ and
$\gamma\neq 0$.

\section{The extended tanh method}
The extended tanh method \cite{abdou} may be described in
the following three steps:\\\\
\noindent \textit{\textbf{Step 1}}. We search exact solutions of
equation (\ref{eq00}) in the form
\begin{equation}\label{eq06}
\begin{cases}
u(x,t)=v(\xi)\\
\xi=x+\lambda t,
\end{cases}
\end{equation}
As a result we have that the equation (\ref{eq00}) is reduced to the
nonlinear ordinary differential equation (ODE) of fifth order
\begin{equation}\label{eq07}
\gamma  v'(\xi ) v(\xi
   )^2+\alpha  v^{(3)}(\xi )
   v(\xi )+\lambda  v'(\xi
   )+\beta  v'(\xi ) v''(\xi
   )+\omega  v^{(5)}(\xi )=0
\end{equation}
\textit{\textbf{Step 2}}.
 We seek solutions of
(\ref{eq07}) in the form
\begin{equation}\label{eq08}
v(\xi)=a_0+\sum_{i=1}^m\left(\,\varphi(\xi)^i+\varphi(\xi)^{-i}\,\right),
\end{equation}
where
\begin{equation}\label{eq09}
\varphi'(\xi)=k+\varphi^2(\xi),
\end{equation}
where $k$ is a parameter to be determined later.\\
 The
Riccati equation (\ref{eq09}) has the general solutions:
\begin{itemize}
\item[\textbf{a)}] If $k<0$ :
$\varphi(\xi)=-\sqrt{-k}\,\tanh(\sqrt{-k}\,\xi)$ and
$\varphi(\xi)=-\sqrt{-k}\,\coth(\sqrt{-k}\,\xi).$
\item[\textbf{a)}] If $k=0$ : $\varphi(\xi)=-\dfrac{1}{\xi}.$
\item[\textbf{c)}] If $k>0$ :
$\varphi(\xi)=\sqrt{k}\,\tan(\sqrt{k}\,\xi)$ and
$\varphi(\xi)=\sqrt{k}\,\cot(\sqrt{k}\,\xi).$
\end{itemize}

Substituting (\ref{eq08}), along with (\ref{eq09}) into (\ref{eq07})
and collecting all terms with the same power in $\varphi(\xi)$,  we
get a polynomial in the variable $\varphi=\varphi(\xi)$. This
polynomial has the form
\begin{equation}\label{eq10}
a\,\varphi(\xi)^{3m+1}+b\,\varphi(\xi)^{2m+3}+c\,\varphi(\xi)^{m+5}+\text{lower
degree terms}
\end{equation}
The parameter $m$ can be found by balancing the high-order linear
term with the nonlinear terms \cite{parker} in  (\ref{eq10}). We
assume that $m\geq 1$ to avoid trivial solutions.  The degrees of
the highest terms are $m+5$ ( the degree of the term
$c\,\varphi(\xi)^{m+5}$ ), $2m+3$ ( the degree of the term
$b\,\varphi(\xi)^{2m+3}$ ) and $3m+1$ ( the degree of the term
$a\,\varphi(\xi)^{3m+1}$ ). The only integer value of $m$ for which
$3m+1=2m+3$ or $3m+1=m+5$ or $2m+3=m+5$ is $m=2$. Equating in
(\ref{eq10}) the coefficients of every power of $\varphi(\xi)$ to
zero, we obtain the following algebraic system in the variables $k$,
$\lambda$, $a_0$, $a_1$, $b_1,\ldots$ :
\begin{itemize}\item $-720 b_2 \omega  k^5-24 b_2^2 \alpha  k^3-12 b_2^2 \beta  k^3-2 b_2^3 \gamma  k=0.$\item $-120 b_2 \omega k^5-30 b_2 b_2 \alpha  k^3-10 b_2 b_2 \beta  k^3-5 b_2 b_2^2 \gamma  k=0.$\item $-1680 b_2 \omega k^4-6 b_2^2 \alpha  k^3-24 a_0 b_2 \alpha  k^3-2 b_2^2 \beta k^3-40 b_2^2 \alpha  k^2-28 b_2^2 \beta  k^2-4 a_0 b_2^2 \gamma  k-4 b_2^2 b_2 \gamma  k-2 b_2^3 \gamma =0.$\item $-240 b_2 \omega  k^4-6 a_0 b_2 \alpha  k^3-24 a_1 b_2 \alpha  k^3+6 a_1 b_2 \beta  k^3-48 b_2 b_2 \alpha  k^2-22 b_2 b_2 \beta  k^2-b_2^3 \gamma  k-3 a_1 b_2^2 \gamma  k-6 a_0 b_2 b_2 \gamma  k-5 b_2 b_2^2 \gamma =0.$\item $-6 a_1 b_2  \alpha  k^3-24 a_2 b_2 \alpha  k^3+2 a_1 b_2 \beta  k^3+8 a_2 b_2 \beta  k^3-1232 b_2 \omega k^3-8 b_2^2 \alpha  k^2-40 a_0 b_2 \alpha  k^2-4 b_2^2 \beta k^2-16 b_2^2 \alpha  k-20 b_2^2 \beta  k-2 a_0 b_2^2 \gamma  k-2 a_2 b_2^2 \gamma  k-2 a_0^2 b_2 \gamma  k-4 a_1 b_2 b_2 \gamma  k-2 b_2 \lambda  k-4 a_0 b_2^2 \gamma -4 b_2^2 b_2 \gamma =0.$\item $-\gamma b_2^3-a_1 k \gamma  b_2^2-6 a_2 k^3 \alpha  b_2-8 a_0 k^2 \alpha  b_2-18 b_2 k \alpha b_2+2 a_2 k^3 \beta  b_2-14 b_2 k \beta  b_2-6 a_0 b_2 \gamma  b_2-a_0^2 k \gamma b_2-2 a_2 b_2 k \gamma b_2-k \lambda  b_2-136  k^3 \omega b_2-38 a_1 b_2 k^2 \alpha +10 a_1 b_2 k^2 \beta -3 a_1 b_2^2 \gamma -2 a_0 a_1 b_2 k \gamma =0.$\item $-2 b_2 \gamma a_0^2-16 b_2 k \alpha  a_0-2 b_2^2 \gamma  a_0-6 a_1 b_2 k^2 \alpha -24 a_2 b_2 k^2 \alpha -2 b_2^2 k \alpha -4 b_2^2 \beta +2 a_1 b_2 k^2 \beta +8 a_2 b_2 k^2 \beta -2 b_2^2 k \beta -2 a_2 b_2^2 \gamma -4 a_1 b_2 b_2 \gamma -2 b_2 \lambda -272 b_2 k^2 \omega =0.$\item $2 a_1 a_2 \beta  k^3+16 a_1 \omega  k^3+2 a_0 a_1 \alpha  k^2+
8 a_2 b_2 \alpha  k^2-2 a_2 b_2 \beta  k^2-16  b_2 \omega\,k^2 -2
a_0 b_2 \alpha  k-8 a_1 b_2 \alpha  k+2 a_1 b_2 \beta k+a_0^2 a_1
\gamma  k+a_1^2 b_2 \gamma  k+2 a_0 a_2 b_2 \gamma  k+2 a_1 a_2 b_2
\gamma  k+a_1 \lambda  k-2 b_2 b_2 \beta -a_1 b_2^2  \gamma -a_0^2
b_2 \gamma -2 a_0 a_1 b_2 \gamma -2 a_2 b_2 b_2 \gamma -b_2 \lambda
=0.$\item $4 a_2^2 \beta  k^3+272 a_2 \omega  k^3+2 a_1^2 \alpha
k^2+16 a_0 a_2 \alpha  k^2+2 a_1^2 \beta  k^2+6 a_1 b_2 \alpha k+24
a_2 b_2 \alpha  k-2 a_1 b_2 \beta  k-8 a_2 b_2 \beta  k+2 a_0 a_1^2
\gamma  k+2 a_0^2 a_2 \gamma  k+4 a_1 a_2 b_2 \gamma  k+2 a_2^2 b_2
\gamma  k+2 a_2 \lambda  k=0.$\item $k \gamma  a_1^3+b_2 \gamma
a_1^2+18 a_2 k^2 \alpha a_1+6 b_2 \alpha  a_1+8 a_0 k \alpha  a_1+14
a_2 k^2 \beta  a_1-2 b_2 \beta a_1+a_0^2 \gamma a_1+2 a_2 b_2 \gamma
a_1+6 a_0 a_2 k \gamma  a_1+\lambda a_1+136 k^2 \omega  a_1+38 a_2
b_2  k \alpha -10 a_2 b_2 k \beta +2 a_0 a_2 b_2 \gamma +3 a_2^2 b_2
k \gamma =0.$\item $2 a_2 \gamma  a_0^2+40 a_2 k \alpha a_0+2 a_1^2
\gamma  a_0+4 a_2^2 k \gamma  a_0+16 a_2^2 k^2 \alpha +6  a_1 b_2
\alpha +24 a_2 b_2 \alpha +8 a_1^2 k \alpha +20 a_2^2 k^2 \beta -2
a_1 b_2 \beta -8 a_2 b_2 \beta +4 a_1^2 k \beta +4 a_1 a_2 b_2
\gamma +2 a_2^2 b_2 \gamma  +4 a_1^2 a_2 k \gamma +2 a_2 \lambda
+1232 a_2 k^2 \omega =0.$\item $\gamma a_1^3+6 a_0 \alpha  a_1+48
a_2 k \alpha  a_1+22 a_2 k \beta  a_1+6 a_0 a_2 \gamma a_1+5 a_2^2 k
\gamma  a_1+240 k \omega  a_1+24  a_2 b_2 \alpha -6 a_2 b_2 \beta +3
a_2^2 b_2 \gamma =0.$\item $2 k \gamma a_2^3+40 k \alpha  a_2^2+28 k
\beta a_2^2+4 a_0 \gamma  a_2^2+24 a_0 \alpha  a_2+4 a_1^2 \gamma
a_2+1680 k \omega  a_2+6 a_1^2 \alpha +2 a_1^2 \beta =0.$\item $5
a_1 \gamma  a_2^2+30 a_1 \alpha a_2+10 a_1 \beta a_2+120 a_1 \omega
=0.$\item $2 \gamma a_2^3+24 \alpha a_2^2+12 \beta a_2^2+720 \omega
a_2=0.$\end{itemize} \noindent\textit{\textbf{Step 3}}. ( This is
the more difficult step ) Solving the previous system for $k$,
$\lambda$, $a_0$, $a_1$, $b_1,\ldots$ , we get $a_1=b_1=0$, so when
$m=2$ solutions have the form
$$v(\xi)=a_0+a_2\varphi^2(\xi)+\dfrac{b_2}{\varphi^2(\xi)}.$$
We get the following
 solutions, where
$$A=2 \alpha +\beta +\sqrt{(2
   \alpha +\beta )^2-40 \gamma\,
    \omega },\quad B=\dfrac{12 \gamma\,  \omega -A
   \beta }{8 \gamma }\quad\text{and}\quad C=\dfrac{(3 A-10 \beta )\, \omega
   }{2 A}:$$

\begin{itemize}

\item $ a_0= -\frac{2 A k}{\gamma }, a_2 = 0, b_0 = -\frac{3 A k^2}{\gamma }, \lambda = 16 B k^2$ :
$$u_1(x,t)= -\frac{A k}{\gamma } \left(2+3 \cot ^2\left(\sqrt{k} \ \left(x+16 B  k^2  t+\right)\right) \right).$$
$$u_2(x,t)=  \frac{A k }{\gamma }\left(1+3 \,\text{csch}^2\left(\sqrt{-k} \left(x+16 B  k^2  t\right)\right) \right).$$

\item $ a_0= -\frac{80 k \omega }{A}, a_2 = 0, b_0 = -\frac{120 k^2\omega }{A}, \lambda = 16 C
k^2$ :
$$ u_3(x,t)= -\frac{40 k \omega }{A} \left(2+3 \cot^2\left(\sqrt{k} \left(x+256 C  k^2 t\right)\right)\right)$$
$$u_4(x,t)= \frac{40 k \omega }{A} \left(
1+3\, \text{csch}^2\left(\sqrt{-k} \left(x+256 C  k^2
t\right)\right) \right)$$

 \item $ a_0= -\frac{2 A k}{\gamma }, a_2 = -\frac{3
A}{\gamma }, b_0 = 0, \lambda = 16 B k^2$ :
$$u_5(x,t)=  -\frac{A
k}{\gamma } \left(2+3 \tan^2\left(\sqrt{k} \left(x+16 B  k^2
t\right)\right) \right)$$
$$u_6(x,t)= -\frac{A k}{\gamma } \left(2-3 \tanh ^2\left(\sqrt{-k}
\left(x+16 B  k^2  t\right)\right) \right).$$

\item $a_0= -\frac{80 k \omega }{A}, a_2 = -\frac{120 \omega }{A}, b_0 = 0,
\lambda = 16 C k^2$ :
$$ u_7(x,t)= -\frac{40 k \omega}{A}  \left(2+3 \tan^2\left(\sqrt{k} \left(x+256 C  k^2 t\right)\right)\right).$$
$$ u_8(x,t)= -\frac{40 k \omega}{A}  \left(2-3 \tanh ^2\left(\sqrt{-k}
\left(x+256 C k^2 t\right)\right)\right).$$

\item $ a_0= -\frac{2 A k}{\gamma }, a_2 = -\frac{3 A}{\gamma }, b_2 =
-\frac{3 A k^2}{\gamma }, \lambda = 256 B k^2$ :
$$ u_9(x,t)= -\frac{A k}{\gamma }
\left(2+ 3 \tan ^2\left(\sqrt{k} \left(x+256 B  k^2
t\right)\right)+3 \cot ^2\left(\sqrt{k} \ \left(x+256 B  k^2
t\right)\right) \right).$$
$$u_{10}(x,t)= -\frac{A k}{\gamma } \left(2- 3 \tanh ^2\left(\sqrt{-k}
\left(x+256 B  k^2 t\right)\right)-3 \coth ^2\left(\sqrt{-k} \
\left(x+256 B  k^2 t\right)\right)  \right).$$

\item $ a_0= -\frac{80 k \omega }{A}, a_2 = -\frac{120 \omega }{A},
b_2 = -\frac{120 k^2 \omega }{A}, \lambda = 256 C k^2$ :
$$u_{11}(x,t)=-\frac{40 k \omega }{A} \left(2+
3 \tan^2\left(\sqrt{k} \left(x+256 C  k^2 t\right)\right)+3
\cot^2\left(\sqrt{k} \left(x+256 Ck^2 t\right)\right) \right).$$
$$u_{12}(x,t)=-\frac{40 k \omega }{A} \left(2 - 3 \tanh ^2\left(\sqrt{-k} \left(x+256 C  k^2 t\right)\right)-3 \coth^2\left(\sqrt{-k} \left(x+256 Ck^2 t\right)\right)
\right).$$
  \end{itemize}

Observe that $A=0$ only if $\omega\,\gamma=0$.
\section{Conclusions}
In this paper, by using the extended tanh method and the help of a
symbolic computation engine, we obtain some exact solutions for the
equation (\ref{eq00}). At the same time, we may find solutions for
\emph{any} particular case of this equation, for example,
(\ref{eq01})-(\ref{eq05}). In our opinion, this result is new in the
literature.

\end{document}